# Crossing the Academic Ocean? Judit Bar-Ilan's Oeuvre on Search Engines studies[1]

**Enrique Orduña-Malea**

Universitat Politécnica de Valencia.
Department of Audiovisual Communication, Documentation and History of Art.
Camino de Vera s/n, Valencia 46022, Spain.
e-mail: enorma@upv.es

**Abstract**
The main objective of this work is to analyse the contributions of Judit Bar-Ilan to the search engines studies. To do this, two complementary approaches have been carried out. First, a systematic literature review of 47 publications authored and co-authored by Judit and devoted to this topic. Second, an interdisciplinarity analysis based on the cited references (publications cited by Judit) and citing documents (publications that cite Judit's work) through Scopus. The systematic literature review unravels an immense amount of search engines studied (43) and indicators measured (especially technical precision, overlap and fluctuation over time). In addition to this, an evolution over the years is detected from descriptive statistical studies towards empirical user studies, with a mixture of quantitative and qualitative methods. Otherwise, the interdisciplinary analysis evidences that a significant portion of Judit's oeuvre was intellectually founded on the computer sciences, achieving a significant, but not exclusively, impact on library and information sciences.

## 1. Introduction

Judit Bar-Ilan was a leading information scientist, with a strong mathematical background, and a final target on users' behaviour, just the academic cocktail I was eager to find at the time when I decided to direct my life towards the Academy. Her influence on my academic training is immeasurable.

Judit sadly passed away on July 16, 2019, and this work aims to pay tribute to her achievements and academic legacy.

Judit received a technical education including B.Sc. in Mathematics and Computer Science–with distinction (1981), M.Sc. in Mathematics–with distinction (1983), and a PhD in Computer Science (1990), all at the Hebrew University of Jerusalem.

After an academic cycle including a postdoctoral position at the Weizmann Institute of Science (1989-90), a visiting Lecturer position at the Department of Mathematics and Computer Science, University of Haifa (1990-91), and being responsible for the seminars in Computer Science at The Open University of Israel (1990-92), she moved back to the Hebrew University of Jerusalem in 1991-92 to become a member of the School of Library and Information Studies, where her academic career–forevermore linked to the Social Sciences–started, first as External Teacher, and later as Teaching Fellow (1992-94), Teacher (1994-98) and Senior Teacher (1998-2002).

---

Later, Judit moved to the Department of Information Science at Bar-Ilan University in 2002, where she was head of Department from 2008 to 2012, and was promoted to Full Professor in 2010.

Judit's outstanding oeuvre comprises over 300 academic publications, including journal articles, book chapters, conference papers, book reviews, to which we must add her teaching dedication and an active role in the community through numerous conference program committee memberships and journal editorial board positions. The impact of Judit's work can be fairly reflected through the nearly 4,000 citations currently received according to Scopus (over 8,000 according to Google Scholar citation profiles).

Judit was active in different fields, such as informetrics and webometrics (search engine studies and link analysis), information retrieval and dynamics, internet research, information behaviour and usability, citation analysis (especially web citation search engines, such as Google Scholar), and altmetrics (Thelwall 2017).

In recognition of her career, Judit was honoured, among other awards, with the Derek de Solla Price Memorial Medal in 2017,[2] awarded by the *International Society for Scientometrics and Informetrics* (ISSI),[3] and with the Research in Information Science Award in 2018, awarded by the *Association for Information Science and Technology* (ASIST).[4]

When I started figuring out the topic for this tribute, I was first tempted to perform a webometric analysis of Judit's personal website[5] or to carry out a content analysis of the results retrieved by Google to the query "judit bar-ilan", following Judit's own footprints in the magnificent tributes and festschrifts she herself had previously paid to Paul Erdos (Bar-Ilan 1998b), Peter Ingwersen (Bar-Ilan 2010) or Eugene Garfield (Bar-Ilan 2018). Then I considered the possibility of performing a bibliometric analysis of Judit's work through Google Scholar or even to go through with an Altmetrics study. All of them were areas in which Judit left her academic mark, and that could faithfully reflect the multidisciplinary impact of her work.

However, while consulting her extensive bibliography, one of her first works published in the journal Scientometrics (Bar-Ilan 1998a), entitled: «On the overlap, the precision and estimated recall of search engines. A case study of the query 'Erdos'», came to my hands. This publication exhibits a large number of quantitative measures applied to several search engines with the purpose of establishing performance evaluation parameters, from an 'informetrics' point of view. This work initiated one of the Judit's main lines of research, and helped, along with the seminal works of Isidro Aguillo, Tomas Almind, Lennart Björneborn, Peter Ingwersen, Mike Thelwall and Liwen Vaughan, among others, to lay the foundations of the so-called Webometrics (i.e., informetrics analyses of the Web).

Search engines studies constitute a large research area, mainly mastered by computer sciences. Scopus indexes currently 22,152 documents (from 1992 to 2019), out of which 15,779 (71.2%) have been published in sources totally or partially classified in this area, while social sciences exhibits just 2,815 contributions (12.7%). One of Judit's main contributions was precisely to study search engines as carriers of information to users, either scholars or general citizens.

[2] https://link.springer.com/article/10.1007%2Fs11192-017-2552-2
[3] http://issi-society.org/awards/derek-de-solla-price-memorial-medal
[4] https://www.asist.org/2018/08/08/bar-ilan-wins-research-award
[5] https://is.biu.ac.il/en/judit

Following this line of though, the first objective of this work is to provide a descriptive and systematic literature review of Judit's contributions dedicated to search engines studies. The second objective is to determine the degree of interdisciplinarity of this specific body of literature, analysing both the cited references (those contributions cited by Judit's work) and the citing documents (those contributions citing Judit's work).

## 2. Method

The first step consisted on identifying the bibliographic corpus dedicated to search engine studies. To do this I accessed to the Judit Bar-Ilan's public profile on Google Scholar Citations,[6] as of 25 December, 2019, which included 230 items.

The selection process was carried out in two consecutive iterations. The first iteration gathered 52 contributions, after reading the title and abstract of each of the 230 items. The second iteration reduced the corpus to a final set of 47 contributions (33 journal articles, 11 conference papers, and 3 book chapters), after a cursory reading of the full text of each pre-selected contribution (See Annex I).

The second step consisted on the realization of a systematic literature review. A detailed reading of the 47 contributions was made in order to extract some basic information, specifically the search engines under study, the research method used, the queries (if any) performed, the number of results analysed (sample size), the date of experiments, and, last but not least, the search engines' parameters and variables studied.

The third step consisted on extracting the cited references from all these contributions. To do this, all cited references from Scopus (42 out of the 47 articles are indexed in this database) were automatically downloaded. The cited references for the remaining five contributions were directly extracted from the manuscripts' full text.

The fourth step consisted on extracting the citing documents. The references of all works citing any of the 47 contributions were automatically downloaded from Scopus.

After this, a data cleansing step (fifth step) was carried out to fix and normalise both cited references and citing sources, due to the significant number of errors encountered.

Finally, the sixth step was dedicated to the interdisciplinarity. In this case, only journal articles were considered, for the sake of clarity.

Each bibliographic reference, either cited reference or citing document, was categorized according to the category assigned to the journal were the article had been published. In order to maintain consistency and coherence, the 27 major thematic categories provided by the *Scopus Subject Areas and Subject Categories* were utilised. When a journal was categorised under more than one major thematic category, a fractional counting (1/n) was used. Therefore, a weighted number of cited references and a weighted number of citations received were obtained.

[6] https://scholar.google.com/citations?user=mkb_14UAAAAJ

This way, a score for each category and contribution was obtained, considering both the articles included in the set of cited references (influential articles for Judit) and the articles included in the set of citing references (articles influenced by Judit). These scores were all transformed into percentage values to minimize size effects.

All process was carried out last week of December, 2019.

# 3. Results

## 3.1. Systematic literature review

Judit initially cultivated this field in a relative lonely way. She was the unique author in 21 out of the 47 selected works. Later, Bluma Peritz (5), Maaya Zhitomirsky-Geffet (5) and specially Mark Levene (14) become her closest collaborators.

The 47 contributions that shape Judit's oeuvre on search engine studies achieve 914 citations according to Scopus. This number climbs to 1,923 in Google Scholar Citation profile.[7] The article entitled «Search engine results over time: A case study on search engine stability» (Bar-Ilan 2003), published in the unfortunately defunct journal Cybermetrics (88 citations according to Scopus; 199 according to Google Scholar), and the article «Data collection methods on the Web for infometric purposes—A review and analysis» (Bar-Ilan 2001), published in Scientometrics (89 citations received computed by Scopus; 180 by Google Scholar), stand out as Judit's most cited contributions on the topic.

Taking apart descriptive and theoretical-oriented documents, 38 contributions out of the 47 provide empirical results on search engines. Annex II contains detailed information about search engines covered, parameters studied, methods employed, queries used, and sample sizes employed. Data collection dates, when available, have also been collected.

Most of Judit's contributions start by acknowledging the Internet as an emerging information medium, where users were experiencing a 'Web document explosion' (Bar-Ilan 1998b). Consequently, Internet in general, and the Web in particular, might become as a potential information and bibliographical source for scientists (Bar-Ilan 2000). Within this ecosystem, search engines appeared to constitute an essential part of the Web (Bar-Ilan 2002). However, Judit's experiments came to demonstrate that the quality and the reliability of most of the available search tools were not satisfactory (Bar-Ilan 2001).

If Annex II (Search engine column) is analysed, one can feel witness to the evolution of the search engine market. Driving through Judit's work we can move from pioneer search engines like Altavista, Excite, Fast, Infoseek, Northern Light or Lycos to the current landscape dominated by Google, including the usage of local search engines (Walla, Morfix, Tapuz, Yandex, Rambler, Voila, Origo-Vizala, etc.) on the route.

[7] A customized profile including the 47 contributions was created for the occasion. Duplicate records were appropriately merged to gather all citations covered by Google Scholar database.

At the end, 43 different search engines were tested, being Google (including different market versions) and Altavista the most widely employed (29 and 18 times respectively).

The review of this body of literature also allows locating beautiful pieces. The rise of Google was prophesied by Judit almost 20 years ago, when she pointed out that "for almost all purposes it will be enough to search Google to get good coverage of a topic on the Internet" (Bar-Ilan 2002). Otherwise, Judit proposed the creation of "vertical search engines and directories per disciplines with high quality control (Bar-Ilan 2001), prophesizing the launch of Google Scholar. The ideal of a search engine serving the scientific community accompanied Judit along different contributions (Bar-Ilan 2005a; 2005b), where even a name 'Webomet', originally coined by Björneborn, was adopted.

All the parameters, variables and indicators used by Judit to characterize and evaluate search engines constitute another essential contribution to the field. Adopting postulates from the Information Retrieval (IR) field, Judit calculated several variables: estimated recall, technical relevance, technical precision, overlap, self-overlap, coverage, relative coverage, and evolution over time. Special attention was paid to the analysis of the stability and fluctuation over time, putting the URLs at the heart of the analyses (lost URLs, dropped URLs, forgotten and totally forgotten URLs, Recovered URLs, etc.).

Following in the wake of Judit's works on search engine studies, we can see a movement from pure informetric methods to content analyses first, and user studies later. From quantitative analyses aimed at discovering the response of search engines as information retrieval systems to characterizing the results offered (content-centred studies) and the user responses (user-centred studies). Judit mixed quantitative and qualitative methods, and gradually she moved from technical precision to ordering results, providing empirical results to the emerging field of search engine optimization (SEO), with users' studies and tailored experimental designs.

The evolution of Judit's works on search engine studies can be observed in the co-occurrence map of keywords included in Figure 1. 'Search engines' (29 occurrences), 'World Wide Web' (14) and 'Information retrieval' (12) stand out as the most used keywords.

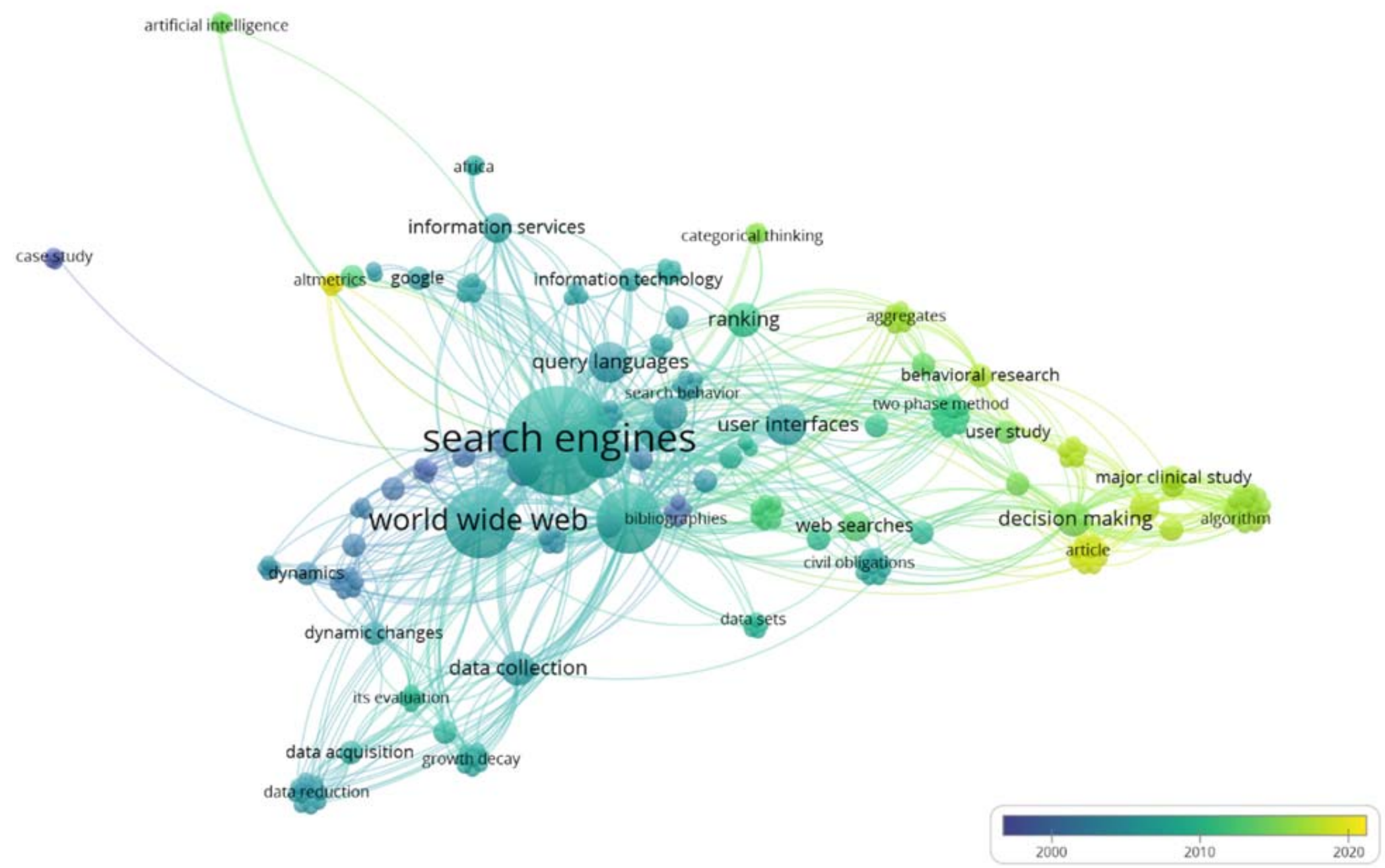

**Figure 1**
**Co-occurrence overlay map of keywords (1998-2019)**
Map generated with WoSViewer. Terms extracted from Scopus database.
Total documents included: 42; Total keywords included: 214.

## 3.2. Interdisciplinarity

Interdisciplinarity remains as a controversial concept in Scientometrics, as nuanced differences between interdisciplinary, multidisciplinary, and cross-disciplinary emerge but remain hard to handle, especially when measured at the journal-level.

As the eminent Albert-Láazló Barabási has recently pointed out in a Twitter thread, whereas 'multidisciplinary' refers to separate disciplines coming together in the same journal, yet remaining distinct,[8] 'interdisciplinarity' refers to integration of disciplines in the same publication. Therefore, Interdisciplinary impact is the diversity of disciplines that a discovery influences, defined by the disciplines that cited the paper.[9] Cross-disciplinarity emerges when a disciplinary paper impacts other disciplines.[10]

Following this terminology, the overall goal of this section is to analyse the interdisciplinary degree of Judit's work on search engine studies.

From the 47 contributions, Judit provided a total of 1,832 cited references, mainly to journal articles (48.5%). However, the great amount of references to online material (24.6%) really stands out. Judit was eclectic and heterodox in her citing profile. She frequently cited newspapers, search engines' webpages with technical information and definitions, reports, dictionaries, encyclopaedias, working papers, discussion lists, conclusions from conference special interest groups, and above all, posts from specialized blogs. Search Engine Watch,[11] a reputed blog devoted to search engine optimization, is cited up to 72 times. Search engine studies are a very highly dynamic area, and the most updated and fresh content is generally found in these online sources.

[8] https://twitter.com/barabasi/status/1193166726663413761
[9] https://twitter.com/barabasi/status/1193166727716245504
[10] https://twitter.com/barabasi/status/1193166730601873408

Conference papers (19.2%) are intensely cited as well, both from computer sciences side (e.g., *International World Wide Web Conference* or *International ACM SIGIR Conference*) and social sciences side (e.g., *ASIS Annual Meeting* or *International Conference of the International Society for Scientometrics and Informetrics*).

Otherwise, the typology of citing sources is obviously more restricted. A total amount of 916 citations have been computed, mainly from journal articles (75.4%) and conference papers (18.4%). Figure 2 shows the distribution of document types according to both cited references and citing sources.

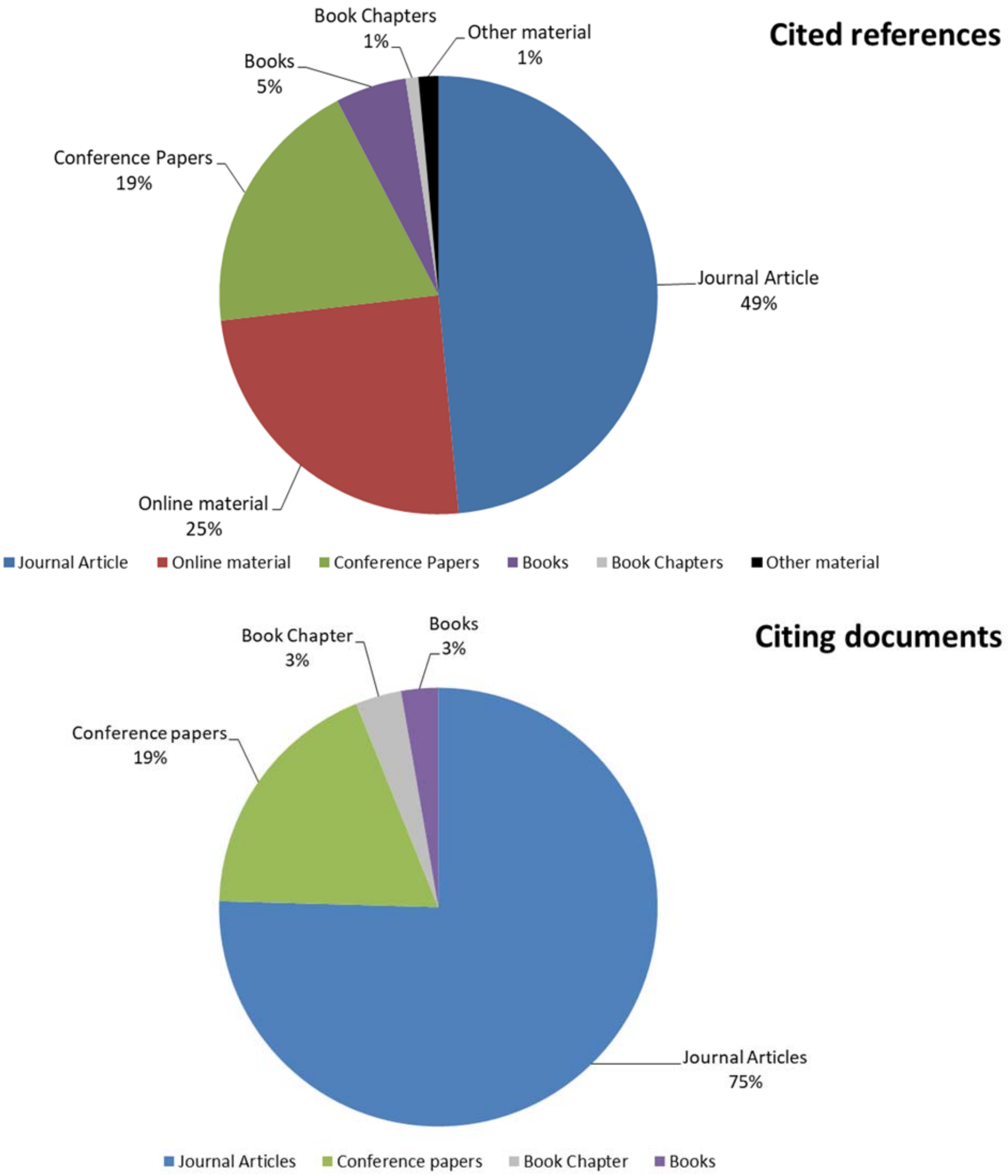

**Figure 2**
**Document types: cited references (up) and citing documents (below)**

[11] https://www.searchenginewatch.com

An author-level analysis have been carried out to reveal those authors most cited by Judit's work on search engines (authors who influenced Judit), and complementarily to this, those authors who most cited Judit (authors influenced by Judit). Table 1 includes the top 20 authors on both sides of the academic coin.

On the one hand we can observe that Judit was influenced to a great extent by authors from computer science, such as Amanda Spink, Bernard Jansen, Clyde Lee Giles, Steve Lawrence or Andrei Broder. It is worth to mention the appearance of David Sullivan (blogger at Search Engine Watch blog) as the second most cited author, as well as the presence of Google as institutional author.

On the other hand, we appreciate a strong influence of Judit's work on authors who, regardless their educational background, have published mainly in the social sciences in general, and webometrics in particular, such as Liwen Vaughan, Isidro Aguillo, Kaivan Kousha or Jose Luis Ortega. In addition we find other important authors with a high technical background like Dirk Lewandowski and Han Woo Park. Finally, Mike Thelwall exhibits a great influence both on the citations received by and provided to Judit.

**Table 1**
**Authors: cited references and citing sources**

| Authors appearing in Cited References | N | Authors appearing in Citing Sources | N |
|---|---|---|---|
| Spink A. | 73 | Thelwall M. | 156 |
| Sullivan D. | 64 | Levene M. | 39 |
| Thelwall M. | 63 | Peritz B.C. | 23 |
| Jansen B.J. | 61 | Vaughan L. | 23 |
| Saracevic T. | 44 | Aguillo I. | 19 |
| Lawrence S. | 44 | Zhitomirsky-Geffet M. | 18 |
| Levene M. | 41 | Kousha K. | 16 |
| Lee Giles C. | 41 | Orduña-Malea E. | 16 |
| Broder A. | 41 | Ortega J.L. | 15 |
| Google | 37 | Wilkinson D. | 14 |
| Kumar R. | 36 | Bhavnani S.K. | 9 |
| Bharat K. | 25 | Park H.W. | 9 |
| Tomkins A. | 23 | Lewandowski D. | 9 |
| Henzinger M. | 23 | Payne N. | 9 |
| Rousseau R. | 22 | Harries G. | 8 |
| Raghavan P. | 22 | Ashman H. | 8 |
| Vaughan L. | 20 | Nelson M.L. | 8 |
| Rajagopalan S. | 20 | Jansen B.J. | 8 |
| Peritz B.C. | 19 | Sud P. | 8 |
| Ingwersen P. | 18 | Schmakeit J.-F. | 8 |

As regards the publication sources, we can observe a similar pattern (Table 2). Taking apart the presence of specialized blogs and conference proceedings, cited references include interdisciplinary journals with a great weight on technical aspects and pure computer sciences journals (e.g., Computer Networks, Lecture Notes in Computer Science). On the other side, the citing documents exhibit a greater presence of journals from library and information sciences. In any case, JASIST, an interdisciplinary journal, appears as the most important source for Judit's works on search engine studies.

**Table 2**
**Authors: cited references and citing sources**

| Journals appearing in Cited References | N | Journals appearing in Citing Sources | N |
|---|---|---|---|
| JASIST[1] | 228 | JASIST[2] | 111 |
| Information Processing & Management | 69 | Scientometrics | 75 |
| Journal of Documentation | 61 | Journal of Information Science | 40 |
| Scientometrics | 60 | Online Information Review | 37 |
| Cybermetrics | 42 | Information Processing & Management | 28 |
| Nature | 33 | Cybermetrics | 20 |
| Journal of Information Science | 30 | ARIST | 18 |
| Online Information Review | 29 | Journal of Informetrics | 17 |
| Computer Networks | 25 | Journal of Documentation | 13 |
| Computer Networks and ISDN Systems | 23 | Library and Information Science Research | 13 |
| Science | 22 | Aslib Journal of Information Management[3] | 20 |
| Lecture Notes in Computer Science | 22 | Journal of Computer-Mediated Communication | 10 |
| Computer | 17 | New Media and Society | 7 |
| Information Retrieval | 15 | International Information and Library Review | 7 |
| Information Research | 13 | Information Research | 7 |
| Interacting with Computers | 10 | Profesional de la Informacion | 6 |
| SIAM Journal on Discrete Mathematics | 9 | Revista Española de Documentación Científica | 6 |
| ACM Transactions on Information Systems | 9 | ACM Transactions on Information Systems | 5 |
| ARIST | 9 | First Monday | 5 |
| Journal of the ACM | 8 | Library Trends | 5 |

[1] Includes Journal of the American Society for Information Science and Technology, Journal of the American Society for Information Science, and Journal of the Association for Information Science and Technology
[2] Includes Journal of the American Society for Information Science and Technology and Journal of the Association for Information Science and Technology
[3] Includes Aslib Journal of Information Management and Aslib Proceedings: New Information Perspectives

If we move towards the thematic categories (only journal articles considered), cited references (n=888 references) are covered both by computer sciences (36.5% of all weighted references) and social sciences (36.3%), followed by decision sciences (11%).

Citing documents (n=688 citations) are concentrated in social sciences (44.2% of all weighted citations received), followed by computer sciences (33.7%) and decision sciences (7.5%). That is, same fields with different percentages (Figure 3). Within social sciences, impact comes mainly from library and information science (478 out of the 568 citations from journals totally or partially categorized under social sciences belong to this subcategory).

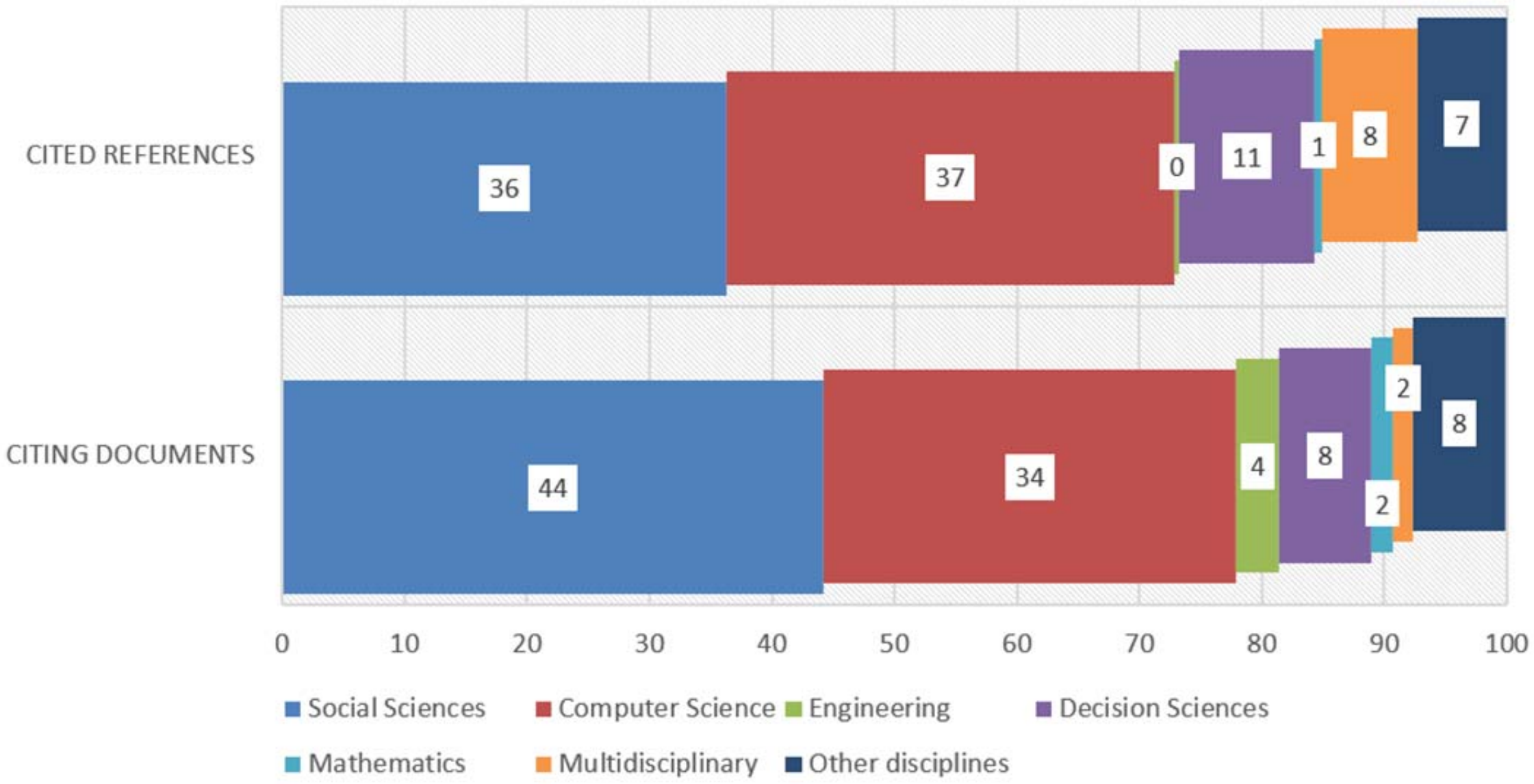

**Figure 3**
**Thematic categories (bibliographic corpus on search engine studies): cited references and citing sources**

Leaving behind the overall behaviour, the performance of particular contributions exhibits interesting information about interdisciplinarity. Figure 4 includes the cited-references/citing-sources balance for two selected contributions (labelled P002 and P025 in Annex I).

P002: This article, originally published in the journal Cybermetrics (Bar-Ilan 2003), was conceived mainly with references from computer science journals (48.1%), but it attracted citations mainly from articles published in social sciences (53.1%).

P025: This article, originally published in the journal Computer Networks (Bar-Ilan, Mat-Hassan and Levene 2006), was conceived with references both from social sciences (28.25%) and computer science journals (25.5%), but it attracted citations mainly from articles published in computer sciences (38.9%), ‘other disciplines’ (20.4%), especially Business, Management and Accounting, and Medicine, and to a lesser extent, social sciences (17.%).

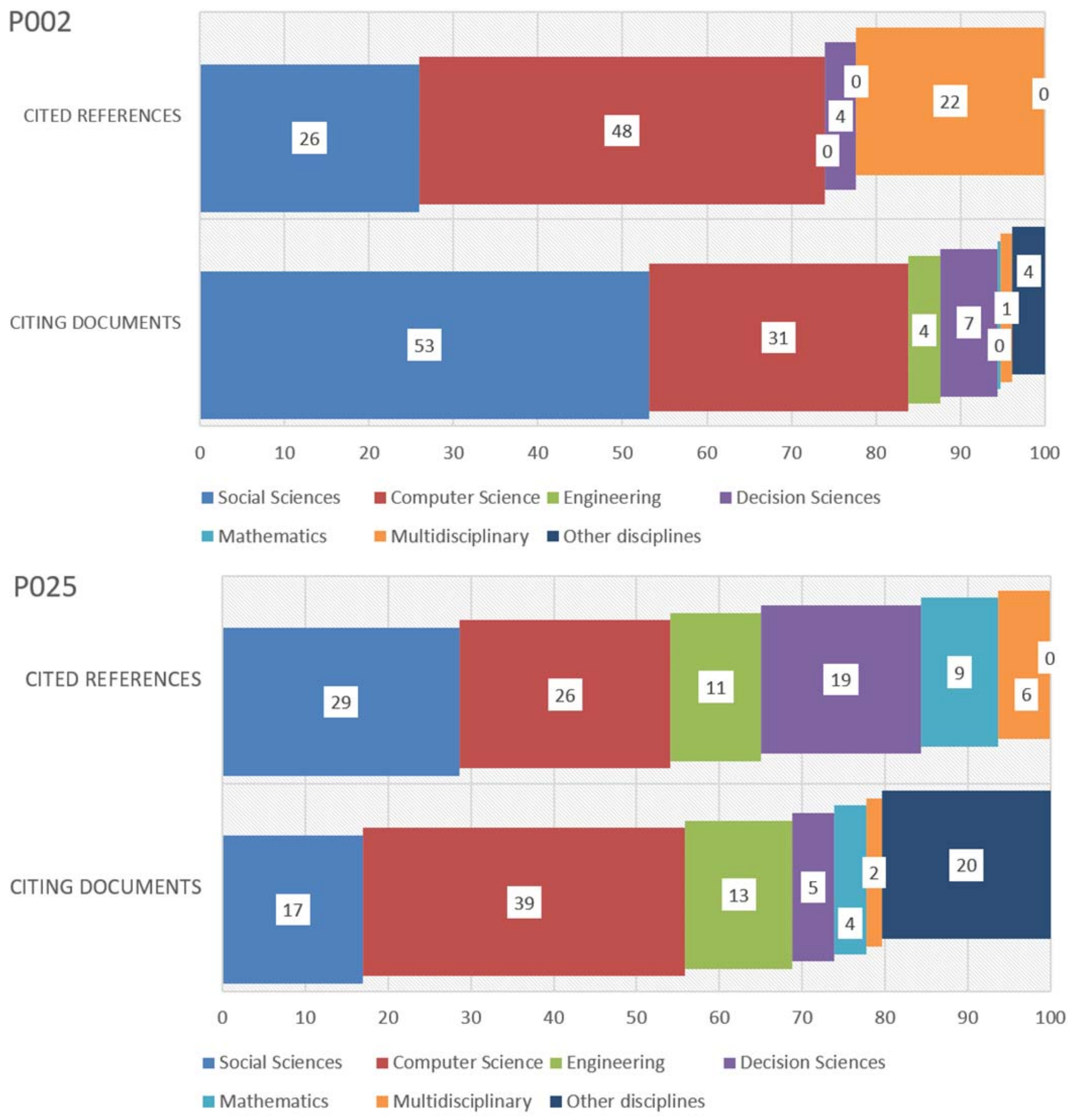

**Figure 4**
**Thematic categories (specific contributions): cited references and citing sources**
(up) P002: Search engine results over time: A case study on search engine stability (Cybermetrics).
(below) P025: Methods for comparing rankings of search engine results (Computer Networks).

To finalize the analysis, we have obtained a two-dimensional coordinates based on the interdisciplinarity of each contribution. To do this, we need to establish a thematic category which will act as a baseline. In this case, the selected category was ‘social sciences’.

For each document, the percentage of cited references outside the social sciences (cited dimension), and the percentage of citing documents outside the social sciences (citing dimensions) were estimated. Then we could plot the coordinates for each of the contributions (Figure 5).

As we can observe, the majority of contributions are located in quadrant 4 (high cited-references interdisciplinarity, high citing-documents interdisciplinarity), with the exception of document P020 (a journal article written in German with only 3 journal articles cited, and 4 citations received), and P018 (a conference paper, which receives just 1 citation from a journal categorized under Social Sciences).

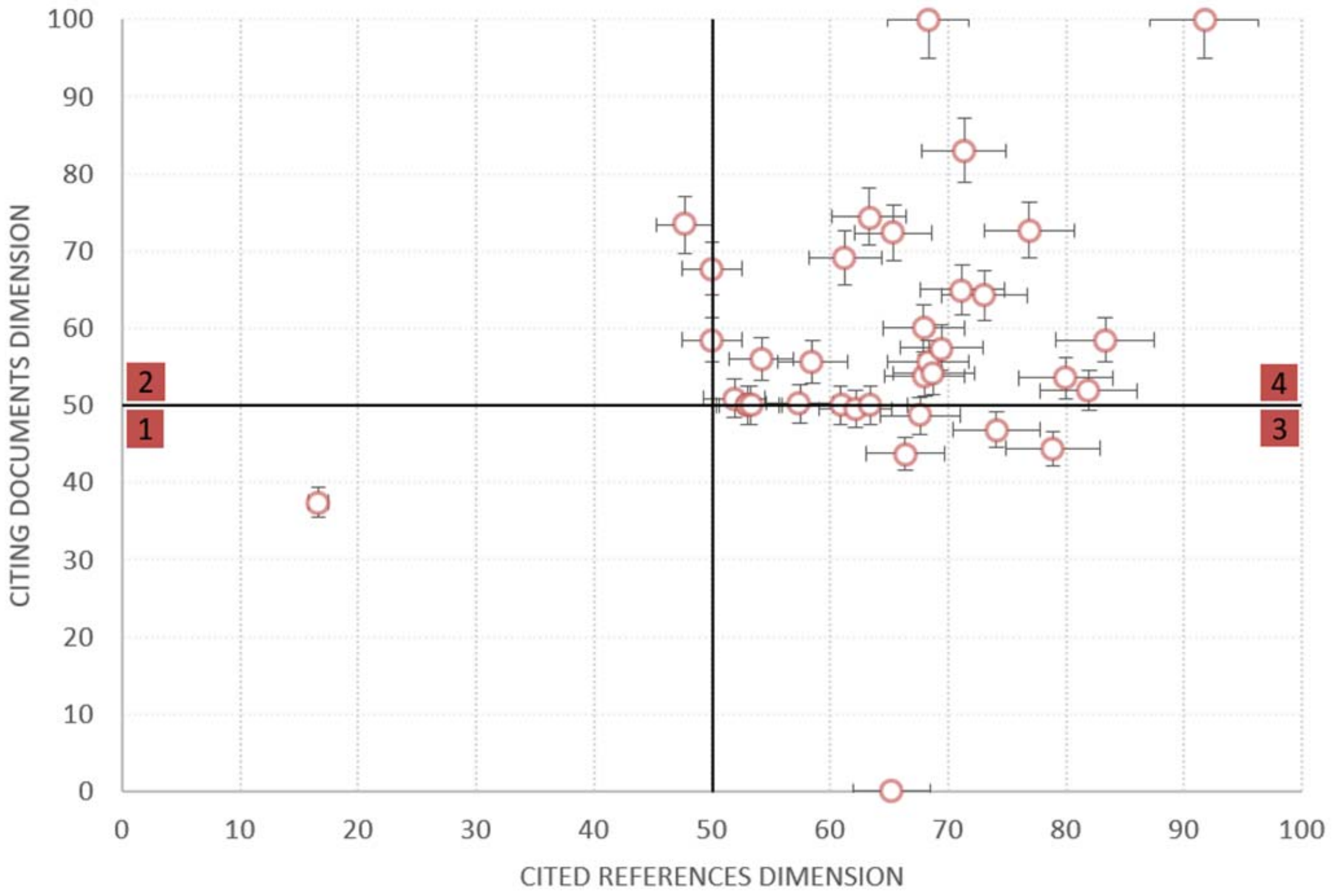

**Figure 5**
**Interdisciplinarity quadrant**

## 4. Discussion and Conclusions

This work reports on the contributions of Judit Bar-Ilan to the search engines studies. To do this, two complementary approaches have been carried out. First, a systematic literature review of 47 publications authored or co-authored by Judit and devoted to this topic. And second, an interdisciplinarity analysis based on the cited references (publications cited by Judit) and citing documents (publications that cite Judit's documents).

The systematic literature review unravels the breadth and depth of Judit's work on search engines, the immense amount of search engines studied and indicators measured. In addition to this, an evolution over the years is detected towards empirical user studies and search engine results rank, with a mixture of quantitative and qualitative methods.

The interdisciplinary analysis shows Judit as a scientist who not only researched the Web but also used it to nurture her publications with numerous mentions of online resources with useful, necessary, updated and rigorous information. That is to say, Judit talked the talk and walked the walk. Otherwise, the results evidence that Judit fed academically on computer sciences, being able to cross the ocean to social sciences, achieving a significant impact especially, but not exclusively, on library and information science.

Throughout this work, we can find some limitations. First, article categorization was performed at the journal-level, which introduces unsurmountable methodological problems. However, recent article-level categorizations still do not solve the problems. Yet, some journal classification inconsistencies were found[12] and manually treated. Expanding the analysis by taking the specific subject categories into account is also advisable. Second, only citing sources indexed in Scopus were considered. Including a wide spectrum of citations received (mainly from Google Scholar) might help to obtain a wider citation scenario. Third, only journal articles were considered in the interdisciplinarity analysis. The inclusion of other document types (mainly book chapters and conference papers) might increase the weight of computer sciences, especially on the citing documents side.

At all events, this work evidences the richness, impact, and interdisciplinarity of Judit's work, and her legacy to the field of search engines studies.

## References

Bar-Ilan, J. (1998a). On the overlap, the precision and estimated recall of search engines. A case study of the query "Erdos." *Scientometrics*, *42*(2), 207–228. https://doi.org/10.1007/bf02458356

Bar-Ilan, J. (1998b). The mathematician, Paul Erdos (1913–1996) in the eyes of the Internet. Scientometrics, *43*(2), 257–267. https://doi.org/10.1007/bf02458410

Bar-Ilan, J. (2000). The web as an information source on informetrics? A content analysis. *Journal of the American Society for Information Science and Technology*, *51*(5), 432–443. https://doi.org/10.1002/(sici)1097-4571(2000)51:5%3C432::aid-asi4%3E3.0.co;2-7

Bar-Ilan, J. (2001). Data collection methods on the web for informetric purposes - A review and analysis. *Scientometrics*, *50*(1), 7–32.

Bar-Ilan, J. (2002). Methods for measuring search engine performance over time. *Journal of the American Society for Information Science and Technology*, *53*(4), 308–319. https://doi.org/10.1002/asi.10047

Bar-Ilan, J. (2003). Search Engine Results over Time-A Case Study on Search Engine Stability. *Cybermetrics*, *2/3*, 1–16. Available at: https://dialnet.unirioja.es/descarga/articulo/1256900.pdf (accessed 15 January 2020).

Bar-Ilan, J. (2005a). Expectations versus reality – Search engine features needed for Web research at mid 2005. *Cybermetrics*, *9*(May), 1–26. Available at https://dialnet.unirioja.es/descarga/articulo/1292632.pdf (accessed 15 January 2020).

Bar-Ilan, J. (2005b). Expectations versus reality - Web search engines at the beginning of 2005. In *Proceedings of ISSI 2005: 10th International Conference of the International Society for Scientometrics and Informetrics* (Vol. 1, pp. 87–96).

Bar-Ilan, J. (2010). The WIF of Peter Ingwersen's website. In B. Larsen, J. W. Schneider, & F. Åström (Eds.), *The Janus Faced Scholar a Festschrift in honour of Peter Ingwersen* (pp. 119–121). Det Informationsvidenskabelige Akademi. Available at: https://vbn.aau.dk/ws/portalfiles/portal/90357690/JanusFacedScholer_Festschrift_PeterIngwersen_2010.pdf#page=122 (accessed 15 January 2020).

Bar-Ilan, J. (2018). Eugene Garfield on the Web in 2001. *Scientometrics*, *114*(2), 389–399. https://doi.org/10.1007/s11192-017-2590-9

Bar-Ilan, J., Mat-Hassan, M., & Levene, M. (2006). Methods for comparing rankings of search engine results. *Computer Networks*, *50*(10), 1448–1463. https://doi.org/10.1016/j.comnet.2005.10.020

Thelwall, M. (2017). Judit Bar-Ilan: information scientist, computer scientist, scientometrician. Scientometrics, *113*(3), 1235–1244. https://doi.org/10.1007/s11192-017-2551-3

[12] Journal of Computer-Mediated Communication: according to Scopus, this journal is categorized under Computer Science. In this work, 'Social sciences' category was added; Plos One: according to Scopus, this journal is categorized under Agricultural and Biological Sciences, Medicine, Biochemistry, Genetics and Molecular Biology. In this work, it was categorized under 'Multidisciplinary'. Science: according to Scopus, this journal is categorized under Multidisciplinary and Arts and Humanities. In this work, only 'Multidisciplinary' was considered.

## Annex I. Bibliographic corpus (n=47 contributions)

| ID | Title | Source | Citations (GS) | Citations (Scopus) | Year |
|---|---|---|---|---|---|
| **p001** | On the overlap, the precision and estimated recall of search engines. A case study of the query "Erdos" | Scientometrics | 51 | 28 | 1998 |
| **p002** | Search engine results over time: A case study on search engine stability. | Cybermetrics | 199 | 88 | 1998 |
| **p003** | The life span of a specific topic on the web: the case of "informetrics": A quantitative analysis | Scientometrics | 50 | 23 | 1999 |
| **p004** | Evaluating the stability of the search tools Hotbot and Snap: a case study | Online information review | 51 | 23 | 2000 |
| **p005** | The Web as an information source on informetrics? A content analysis | JASIS | 82 | 39 | 2000 |
| **p006** | Data collection methods on the Web for infometric purposes—A review and analysis | Scientometrics | 180 | 89 | 2001 |
| **p007** | How much information do search engines disclose on the links to a web page? A longitudinal case study of the 'cybermetrics' home page | Journal of information science | 44 | 19 | 2002 |
| **p008** | Criteria for Evaluating Information Retrieval Systems in Highly Dynamic Environments. | CEUR Workshop Proceedings | 7 | 0 | 2002 |
| **p009** | Methods for measuring search engine performance over time | JASIST | 117 | 52 | 2002 |
| **p010** | How do search engines handle non-English queries?-A case study. | WWW (Alternate Paper Tracks) | 29 | | 2003 |
| **p011** | Evolution, continuity, and disappearance of documents on a specific topic on the web: A longitudinal study of "informetrics" | JASIST | 79 | 50 | 2004 |
| **p012** | Dynamics of Search Engine Rankings-A Case Study. | WebDyn@ WWW | 14 | 2 | 2004 |
| **p013** | Search engine ability to cope with the changing web | Web dynamics | 32 | | 2004 |
| **p014** | The use of Web search engines in information science research | Annual Review of Information Science and Technology (ARIST) | 136 | 71 | 2004 |
| **p015** | Comparing rankings of search results on the web | Information Processing & Management | 109 | 43 | 2005 |
| **p016** | From the search problem through query formulation to results on the web | Online Information Review | 25 | 8 | 2005 |
| **p017** | How do search engines respond to some non-English queries? | Journal of Information Science | 75 | 38 | 2005 |
| **p018** | Expectations Versus Reality–Web Search Engines at the Beginning of 2005 | Proceedings of ISSI 2005 | 2 | 1 | 2005 |
| **p019** | Expectations versus reality–Search engine features needed for Web research at mind | Cybermetrics | 61 | 31 | 2005 |
| **p020** | Tauglichkeit von Suchmaschinen für deutschsprachige Abfragen: Schwerpunktthema Suchmaschinen | Information-Wissenschaft und Praxis | 7 | 4 | 2005 |
| **p021** | Mark Levene An Introduction to Search Engines and Web Navigation. Addison Wesley, Pearson Education (2006). ISBN 0-321-30677-5.£ 39.99. 365 pp. Softbound | The Computer Journal | 0 | | 2006 |
| **p022** | Methods for evaluating dynamic changes in search engine rankings: a case study | Journal of Documentation | 17 | 9 | 2006 |
| **p023** | Web links and search engine ranking: The case of Google and the query "jew" | JASIST | 25 | 18 | 2006 |
| **p024** | False Web memories: A case study on finding information about Andrei Broder | First Monday | 5 | 3 | 2006 |
| **p025** | Methods for comparing rankings of search engine results | Computer networks | 161 | 82 | 2006 |
| **p026** | Analysis of queries reaching SHIL on the web–an information system providing citizen information | International Workshop on Next Generation Information Technologies and Systems | 0 | 0 | 2006 |
| **p027** | Popularity and findability: Log analysis of search terms and queries for public services | ILAIS 2006 Conference | 0 | | 2006 |
| **p028** | Position paper: Access to query logs—an academic researcher's point of view | Query Log Analysis Workshop, WWW | 25 | | 2007 |
| **p031** | Manipulating search engine algorithms: the case of Google | Journal of Information, Communication and Ethics in Society | 26 | 13 | 2007 |
| **p032** | Popularity and findability through log analysis of search terms and queries: the case of a multilingual public service website | Journal of Information Science | 25 | 14 | 2007 |
| **p033** | User rankings of search engine results | JASIST | 66 | 42 | 2007 |
| **p034** | The lifespan of "informetrics" on the Web: An eight year study (1998–2006) | Proceedings of ISSI 2007 | | 0 | 2007 |
| **p036** | The lifespan of "informetrics" on the Web: An eight year | Scientometrics | 49 | 25 | 2009 |
| | study (1998–2006) | | | | |
| **p037** | A method for measuring the evolution of a topic on the Web: The case of "informetrics" | JASIST | 18 | 13 | 2009 |
| **p038** | Topic-specific analysis of search queries | Proceedings of the 2009 workshop on Web Search Click Data | 22 | 8 | 2009 |
| **p039** | Users' views on country-specific search engine results | Proceedings of the ASIST | 0 | 0 | 2009 |
| **p040** | Presentation bias is significant in determining user preference for search results—A user study | JASIST | 77 | 46 | 2009 |
| **p041** | A method to assess search engine results | Online Information Review | 16 | 9 | 2011 |
| **p042** | The impact of task phrasing on the choice of search keywords and on the search process and success | JASIST | 24 | 11 | 2012 |
| **p043** | Search Engines and Hebrew-Revisited | Language, Culture, Computation. Computing-Theory and Technology | 0 | 0 | 2014 |
| **p045** | How and why do users change their assessment of search results over time? | Proceedings of the ASIST | 4 | 1 | 2015 |
| **p046** | Testing the stability of "wisdom of crowds" judgments of search results over time and their similarity with the search engine rankings | Aslib Journal of Information Management | 6 | 4 | 2016 |
| **p048** | A Markov chain model for changes in users' assessment of search results | PloS one | 3 | 3 | 2016 |
| **p049** | Analysis of change in users' assessment of search results over time | JASIST | 3 | 3 | 2017 |
| **p050** | Categorical relevance judgment | JASIST | 1 | 1 | 2018 |
| **p051** | Eugene Garfield on the Web in 2001 | Scientometrics | 0 | 0 | 2018 |
| **p052** | Data Collection from the Web for Informetric Purposes | Springer Handbook of Science and Technology Indicators | 0 | 0 | 2019 |

Note: missing numbers (P29, P30, P35, P44, and P47) correspond with documents excluded during the second iteration of the selection process.

## Annex II. Systematic analysis: indicators measured, methods employed, search engines covered, queries analysed and sample sizes.

| Article ID | Indicators measured | Method | Search engine | Queries analysed | Sample | Rounds |
|---|---|---|---|---|---|---|
| P001 | Precision; Technical precision; Estimated recall; Overlap; Coverage; Evolution | Informetrics | Altavista; Excite; Infoseek; Lycos; Magellan; Opentext | 1 query:<br>Erdos | 6,681 URLs | 6 Rounds.<br>Monthly.<br>Nov 1996<br>to Dec 1997 |
| | Coverage; Overlap | Informetrics | Altavista; Excite; Hotbot; Infoseek; Lycos; OpenText | 1 query:<br>Bibliometrics AND growth | 146 URLs | |
| P002 | Coverage; Evolution; Relative coverage; Total relative coverage; Technical precision; Technical relevance; Fluctuation (URL Recovery; URL Permanence); Self-Overlap | Informetrics | Altavista; Excite; Hotbot; Infoseek; Lycos; Northern Light | 1 query:<br>informetrics OR informetric | 1,268 URLs | 5 Rounds.<br>Monthly.<br>Jan to Jun 1998 |
| P003 | Fluctuation; Change type (minor and considerable); Change stability (stagnant and dynamic) | Content Analysis<br>Informetrics | Altavista; Excite; Hotbot; Infoseek; Lycos; Northern Light | 1 query:<br>informetrics or informetric | 1,268 URLs | 6 Rounds.<br>Monthly.<br>Jan to Jun 1998 |
| P004 | Coverage; Query size; Query type; Technical precision; Fluctuation (lost URLs, recovered URLs, Dropped URLs) | Informetrics | Hotbot; Snap's Power Search | 20 queries:<br>WebFerretPro; last total eclipse of the Millenium; ``Erich Segal'' +<br>Doctors; ``existential therapy'' AND NOT (anxiety OR psychotherapy); http://sites.huji.ac.il/IFLA2000/66intro.htm; protochlorophyllide; Colima Volcano; onomatopoeia + Japanese; non-repudiation AND NOT (privacy OR security); http://www.altavizsla.matav.hu; caprylic; Lawrence Olivier; ``Six Day War'' + Golan; (``chinese noodles'' OR ``chinese fried rice'') AND NOT pork; http://www.neci.nj.nec.com/homepages/lawrence/; Nabucco; Charlie Daniels Band; Teletubbies + Dipsy + ``Tinky Winky''; (``citation analysis'' OR ``co-citation analysis'') AND NOT ISI; http://www.huji.ac.il | NA | Daily.<br>Sep to Oct 1999. |
| P005 | Coverage; Precision; Multiplicity; Recall | Content Analysis | Altavista; Excite; Hotbot; Infoseek; Lycos; Northern Light | 1 query:<br>Informetrics OR informetric | 942 URLs | 1 Round.<br>Jun 1998 |
| P006 | Coverage | Informetrics | Altavista; Northern Light; Hotbot; Fast | 8 queries:<br>ccTLD: .br; .nl<br>gTLD:.com, .edu, .org, .gov, .net and .mil). | NA | 1 Round.<br>2 Sep 2000 |
| | | | Altavista<br>Northern Light | 3 queries:<br>industry AND government.; university AND government.; university AND industry AND government | | |
| | | | Altavista; Northern Light | 2 queries:<br>"University" (Netherlands)<br>"Industry" (Netherlands | | |
| | Coverage; Relevance; Self-Overlap | | Google; Webtop; Altavista; Fast; Northern Light; Iwon; Snap | 1 query:<br>Webometrics | 308 URLs | |
| P007 | Coverage (link pages; concealed pages); Technical Precision | Content Analysis<br>Informetrics | Altavista; Raging Search; Fast; Google; Hotbot; Iwon; Northern Light | 1 LINK DOMAIN query per search engine:<br>link:www.cindoc.csic.es/cybermetrics/cybermetrics.html<br><br>Several LINK URL queries like<br>url:www.aaa.bbb/ccc.htm | 456 total URLs | 4 Rounds.<br>Jan 2001 to Jan 2002 |
| P009 | Coverage; Relative coverage; Technical Precision; Fluctuation; Self-overlap | Informetrics | Altavista; Excite; Fast; Hotbot; Google; Northern Light | 1 query:<br>aporocactus | NA | 33 Rounds.<br>Weekly and Monthly.<br>Jan 2000 to Jan 2001. |
| P010;<br>P017 | Coverage | Informetrics | Yandex; Rambler; Aport | 9 queries in Russian:<br>Окно; Окон; белый; Белый; человек шел; люди идут; люди идут; начинать; начать | NA | 1 Round.<br>Nov 2002 |
| | | | Voila; AOL France; La Toile | 5 queries in French<br>Electricite ; électricité ; l'électricité; cheval; chevaux | | |
| | | | Origo-Vizsla; Startlap; Heureka | 8 queries in Hungarian<br>Kar; kár; kutya; kutyák; falu; falvak; javítás; kijavítás | | |
| | | | Morfix; Walla | 8 queries in Hebrew<br>[universita]; [hauniversita]; [bauniversita]; [universitat]; [veshehauniversita]; [mehabait]; [bait]; [midbar/medaber/midavar] | | |
| | | | Altavista; Fast; Google | 30 queries (in each of the languages) | | |
| P011 | Coverage; Growth rate (evolution); Fluctuation (URL Modification, URL Disappearance, URL Persistence) | Content Analysis | Altavista; Excite; Hotbot; Infoseek; Lycos; Northern Ligh; Fast; Google; Teoma; Wisenut | 1 query:<br>Informetrics OR informetric | 7,063 URLs | 4 Rounds. Yearly.<br>1998, 1999, 2002, 2003 |
| P012 | Coverage evolution; Overlap; Self-overlap; Results rank | Informetrics | Google.com; Google.co.uk;; Google.co.il; Alltheweb | 10 queries<br>Modern architecture; Web data mining; World rugby; Web personalization; Human cloning; Internet security; Organic food; Snowboarding; DNA evidence; Internet advertising techniques | 27 users | 2 Rounds. Twice a day.<br>Oct 2003 to Jan 2004 |
| P015 | Rank overlap | Informetrics | Google; Alltheweb; Altavista; Hotbot | 15 queries | 16,985 URLs | 1 Round.<br>Dec 2003 |
| P016 | Search instructions; query formulation | User study | No specific search engine | 178 queries | 35 users | 1 Round.<br>May 2003 |
| P018;<br>P019 | Domain Coverage | Informetrics | Google; Yahoo; MSN Beta | 4 queries:<br>ccTLD: .hu; .ca; .dj; .sr | NA | 1 Round.<br>Jan 2005 |
| P022 | Overlap; Self-overlap; Results rank; Change average ranking | Informetrics | Google; Alltheweb | Same Record P012. | NA | 2 Rounds. Twice a day.<br>Oct 2003; Jan 2004 |
| P023 | Link page characteristics; Link characteristic; Rank position; Link features | Content Analysis | Google | 1query:<br>'jew' | Site1: 689 pages<br>Site2: 294 pages | 1 Round.<br>Aug 2004 |
| P024 | Search tasks | User study | Google; Altavista; Alltheweb; Teoma; Yahoo; MSN | 2 queries:<br>andrei broder<br>andrei broder bio | 49 participants<br>1 page | 1 Round.<br>May 2005 |
| P025 | Overlap; Self-overlap; Rank variability | Informetrics | Google; Yahoo; Teoma; Google Images; Yahoo images; Picsearch | 5 queries<br>US elections 2004; DNA evidence; Organic food; Twin towers; Bondi beach | NA | 2 Rounds. Once a day.<br>Nov2004; Feb 2005 |
| P026;<br>P027;<br>P032 | Query syntax; Query frequency; Query length; Query output; Query evolution; Queries from search engines | Content analysis<br>Web-log analysis | No search engine | 266,295 queries | 1 site:<br>http://shil.info | 1 Round.<br>Mar 2005 to Oct 2005 |
| P033 | Ranking overlap; User ranking; USER – SE Similarity; Popularity; Relative relevance | Informetrics<br>User study | Google; MSN; Yahoo | 12 queries<br>'search engine coverage'; Glycemix index; "web preservation"; Genetic engineering; Stop smoking; Blood test<br>Indexing; Semantic web; Bird flu; Ranking metasearch; Atkins diet | 67 participants<br>120 results | 3 week long round.<br>Nov 9 to 29, 2005 |
| P036 | Coverage; Coverage evolution; URL persistence | Content Analysis<br>Informetrics | Altavista; Excite; Hotbot; Infoseek; Lycos; Northern Light; Google; Teoma; Wisenut; Gigablast; Yahoo; Exalead; MSN | 4 queries :<br>Informetrics or informetric ; informetrics-scientometrics ; informetrics scientometrics ; informetrics site:.es –filetype:pdf | 36,282 URLs | 7 Rounds. Yearly.<br>1998; 1999, 2002, 2003, 2004, 2005, 2006 |
| P037 | Technical relevance<br>URL intermittence; URL lost; URL forgot; URL recovered | Informetrics | Altavista; Excite; Hotbot; Infoseek; Lycos; Northern Light; Google; Teoma; Wisenut; Gigablast; Yahoo; Exalead; MSN | 1 query:<br>Informetrics or informetric | NA | 7 Rounds. Yearly.<br>1998; 1999, 2002, 2003, 2004, 2005, 2006 |
| P039;<br>P041 | Ranking Overlap; User Ranking overlap; SE-User similarities | User study | Google (Google.com Google.co.uk Google.co.il)<br>Live Search (live.com; UK search; Israel search) | 9 queries:<br>[Social Networks facebook]; [Hilary Clinton]; BMI; Israel; [Skin cancer prevention]; [html for beginners]; [Olimpics Beijing]; [World Health Organization]; [Google new developments] | 283 total URLs<br>24 users | 2 stages.<br>July 2008 |
| P040 | Rank order user preference | User study<br>Questionnaire | Google; Windows Live; Yahoo | 13 queries:<br>Anthrax; Making money on the internet; Plasma vs. LCD; Prague tourist sights; Rembrandt; Ronaldinho; Calculating Page Rank; Search optimization; Free antispyware; Sudoku; Andrei Broder; Louvre map | 120 results<br>65 users | 1 Round.<br>October 2006 |
| P043 | Coverage; Freshness | Informetrics | Google (google.co.il); Walla; Morfix; MSN; Tapuz; Yahoo | 15 queries:<br>[university]; [universities]; [The university]; [to the university]; [in the university]; [from the university]; [The university OR of the university OR in the university]; [University OR universities OR the university OR to the university OR in the university OR from the university OR university of]; [Library] two spelling variants; [recipes]; [recipe]; [the recipes]; [cellphones]; [cellphone] two spelling variants; [Western Galilee College] two spelling variants | NA | 1 Round.<br>July 2007 |
| P042 | Search tasks | Questionnaire<br>Log files<br>User study | Google | 4 tasks:<br>Task Online Spending; Task Financial concern; Task Children; Task bank | 100 users<br>88 log files | 1 Round.<br>Jun to Jul 2007 |
| P045 | User ranking relevance | User study | Google | 1 query:<br>“cyber warfare” | 20 results<br>35 individuals | 3 Rounds.<br>n.d. |
| P046;<br>P049 | User ranking relevance; User ranking relevance change; URL rank; User-SE rank overlap; Coarseness; Locality | User study | Google<br>Bing | 2 queries:<br>Big data<br>[Alzheimer] in hebrew | 20 URLs per query<br>87 users | 2 Rounds.<br>n.d. |
| P048 | Rank relevance change | User study | Google<br>Bing | 3 queries:<br>Big data<br>[Alzheimer] in hebrew<br>“cyber warfare” | 120 users | 2-3 Rounds.<br>n.d. |
| P049 | Category-based relevance; Average concordance; Swap ratio | User study | Google | 2 queries:<br>Atkins diet<br>Cloud computing | Sets of 20 results<br>86 users | 3 Rounds.<br>n.d. |
| P051 | Coverage; Link pages categorization | Content Analysis | Altavista; Fast; Google Hotbot; Northern Light | 5 queries:<br>‘Eugene Garfield’; ‘Garfield Eugene’; ‘Gene Garfield’; ‘E. Garfield’; ‘Garfield E’ | 4120 URLs gathered<br>1073 URLs analysed | 1 Round.<br>August 2011 |
| P052 | Coverage | Informetrics | Google; Bing; Yahoo | 26 queries:<br>gTLP: .com; .org; .edu; .net, .gov; .mil<br>ccTLP: .uk, .ca.; .au; .nz. ; .es; .fr; .de; .il ; .cn ; .ru ; .br ; .za<br>Yahoo Altavista; Yahoo AND Altavista; Altavista Yahoo; Altavista AND Yahoo; Altavista; Yahoo; Altavista OR Yahoo; Altmetrics | NA | 1 Round.<br>December 2017 |

Notes: [query]: Queries in Hebrew; NA: data not applicable or available; NA: No Data Available